\documentclass[aps,pra,twocolumn,superscriptaddress]{revtex4}
\usepackage{graphicx,amsmath,amssymb,amsthm,enumerate,color}

\begin{document}
\title{Capacity of optical reading, Part 1: Reading boundless error-free bits\\ using a single photon}
\author{Saikat Guha}
\affiliation{Quantum Information Processing (QuIP) Group, Raytheon BBN Technologies, Cambridge, Massachusetts 02138, USA}
\author{Jeffrey H.~Shapiro} \affiliation{Research Laboratory of Electronics, Massachusetts Institute of Technology, Cambridge, Massachusetts 02139, USA}

\begin{abstract}
We address the problem of how efficiently information can be encoded into and read out reliably from a passive reflective surface that encodes classical data by modulating the amplitude and phase of incident light. We show that nature imposes no fundamental upper limit to the number of bits per that can be read per expended probe photon, and demonstrate the quantum-information-theoretic trade-offs between the photon efficiency (bits per photon) and the encoding efficiency (bits per pixel) of optical reading. We show that with a coherent-state (ideal laser) source, an {\em on-off} (amplitude-modulation) pixel encoding, and shot-noise-limited direct detection (an overly-optimistic model for commercial CD/DVD drives), the highest photon efficiency achievable in principle is about 0.5\,bits read per transmitted photon. We then show that a coherent-state probe can read unlimited bits per photon when the receiver is allowed to make joint (inseparable) measurements on the reflected light from a large block of phase-modulated memory pixels. Finally, we show an example of a spatially-entangled non-classical light probe and a receiver design---constructible using a single-photon source, beam splitters, and single-photon detectors---that can in principle read any number of {\em error-free} bits of information. The probe is a single photon prepared in a uniform coherent superposition of multiple orthogonal spatial modes, i.e., a W-state. The code and joint-detection receiver complexity required by a coherent-state transmitter to achieve comparable photon efficiency performance is shown to be {\em much} higher in comparison to that required by the W-state transceiver, although this advantage rapidly disappears with increasing loss in the system.  
\end{abstract}
\maketitle

\section{Introduction}

\begin{figure}
\centering
\includegraphics[width=\columnwidth]{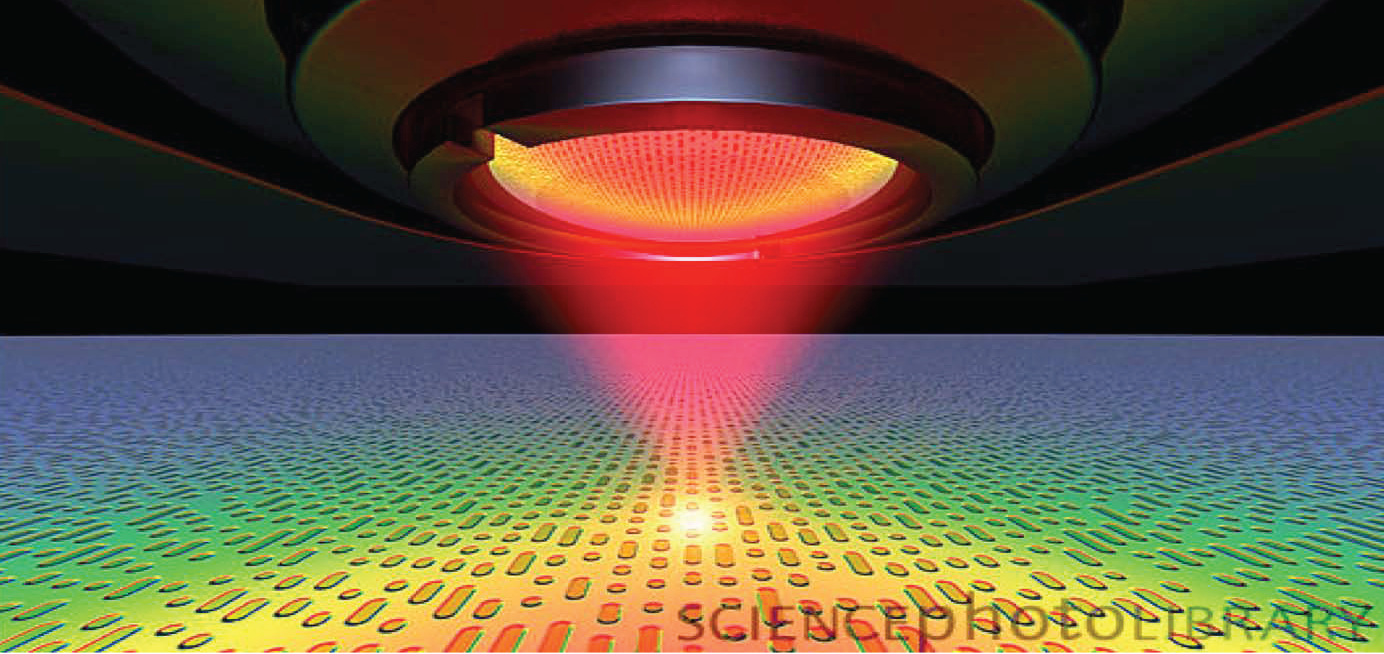}
\caption{Artist's impression of the CD drive's read laser shining on the surface of an optical disc [picture courtesy: Science Photo Library].}
\label{fig:CDmodel_a}
\end{figure}

Optical discs, such as CDs and DVDs, are ubiquitous. The surface of the CD contains a long spiral track of data, along which there are flat reflective areas called {\em land} and non-reflective {\em bumps} (see Fig.~\ref{fig:CDmodel_a}), representing binary $1$ and binary $0$, respectively. The drive shines a laser at the surface of the CD to read data. The detector photocurrent tracks the intensity of the reflected light, which the drive converts into estimates of $1$s and $0$s. There is an extensive literature and ongoing research on evaluation of information-theoretic capacities of optical storage, error-correcting codes, and techniques to make the storage and readout more efficient~\cite{Cen06, Car10}.  The majority of that work, however, concentrates on what can be achieved by optimizing existing technology, as opposed to establishing what are the true ultimate limits---imposed by the laws of quantum mechanics---on optical reading of information that has been encoded into a passive reflecting medium.  

Fundamentally, the performance of any optical communication or imaging system is limited by noise of quantum-mechanical origin, and optical reading of information is no exception. In order to delineate the ultimate performance of optical reading limited only by the laws of physics, an analysis within a full quantum-mechanical framework must therefore be done. Some examples of early work relevant to quantum reading include Shapiro's {\em number-product vacuum} states for zero-error reading of phase-conjugate-encoded pixels~\cite{Sha93}, Acin's work on optimally distinguishing two unitary transformations~\cite{Aci01}, and D'Ariano {\em et al.}'s demonstration that entanglement can improve the precision of estimating an unknown transformation~\cite{DAr01}. In a suite of recent work by Pirandola and others~\cite{Pir11, Pir11a, Osa11, Inv11, Nai11, Nai11a, Nai12, Dall12}, it has been shown that non-classical light paired with non-standard detection techniques can read data more reliably than can a coherent-state (laser) probe, i.e., at a given transmitted-photon budget the former can discriminate between a set of reflectivity-phase values for a pixel with a lower probability of error than the latter.

Lower error probability in discriminating signals from a modulation constellation does not automatically translate to increased {\em capacity}, i.e., the sustained reliable rate of reading that is achievable with an optimal modulation, code, and receiver. Attaining the quantum-limited capacity---the {\em Holevo limit}~\cite{Hol79,Hau96}---requires joint-detection receivers (JDRs), whenever the modulation constellation's quantum states are not mutually orthogonal. JDRs  make collective measurements on the reflected light from many memory pixels---which cannot be realized by detecting the reflected light from each pixel individually---followed by optimal joint post-processing of the classical measurement outcomes~\cite{Guh11a}. Recent work on the capacity of optical reading~\cite{Pir11a} evaluated achievable rates that employ JDRs to detect codewords constructed from binary-amplitude pixel modulation.  These achievable rates, however, fall significantly short---in both the capacity and photon efficiency of optical reading---when amplitude \em and\/\rm\ phase modulation are taken into account.

In this paper, we address the following fundamental question. What is the ultimate upper limit to the number of bits of information that can be reliably read using an optical probe with a given mean photon-number budget when information is encoded using a reflective surface that can passively modulate a combination of the amplitude and phase of the probe light~\cite{Guh11}? We show that there is no upper limit to the number of bits that can be read reliably per expended photon. We also show that with a coherent-state source, an on-off pixel modulation, and ideal direct detection (an overly-optimistic model for commercial CD/DVD drives), the highest photon information efficiency (PIE) achievable is about 0.5\,bits per transmitted photon. We then show that a coherent-state probe can read boundless bits per transmitted photon when the receiver is allowed to make joint measurements on the reflected light from a large block of phase-modulated memory pixels.  We show one structured design for such a JDR that can attain an unbounded PIE with a coherent-state transmitter, which is not possible using any conventional optical transceiver. Finally, we show an example of a spatially-entangled non-classical optical probe and an explicit receiver design---constructible using a single-photon source, beam splitters, and single-photon detectors---that in principle can read any number of {\em error-free} bits of information using a single transmitted photon. The probe is a single photon in a uniform superposition of multiple spatial modes, viz., a W-state. 
\begin{figure}
\centering
\includegraphics[width=\columnwidth]{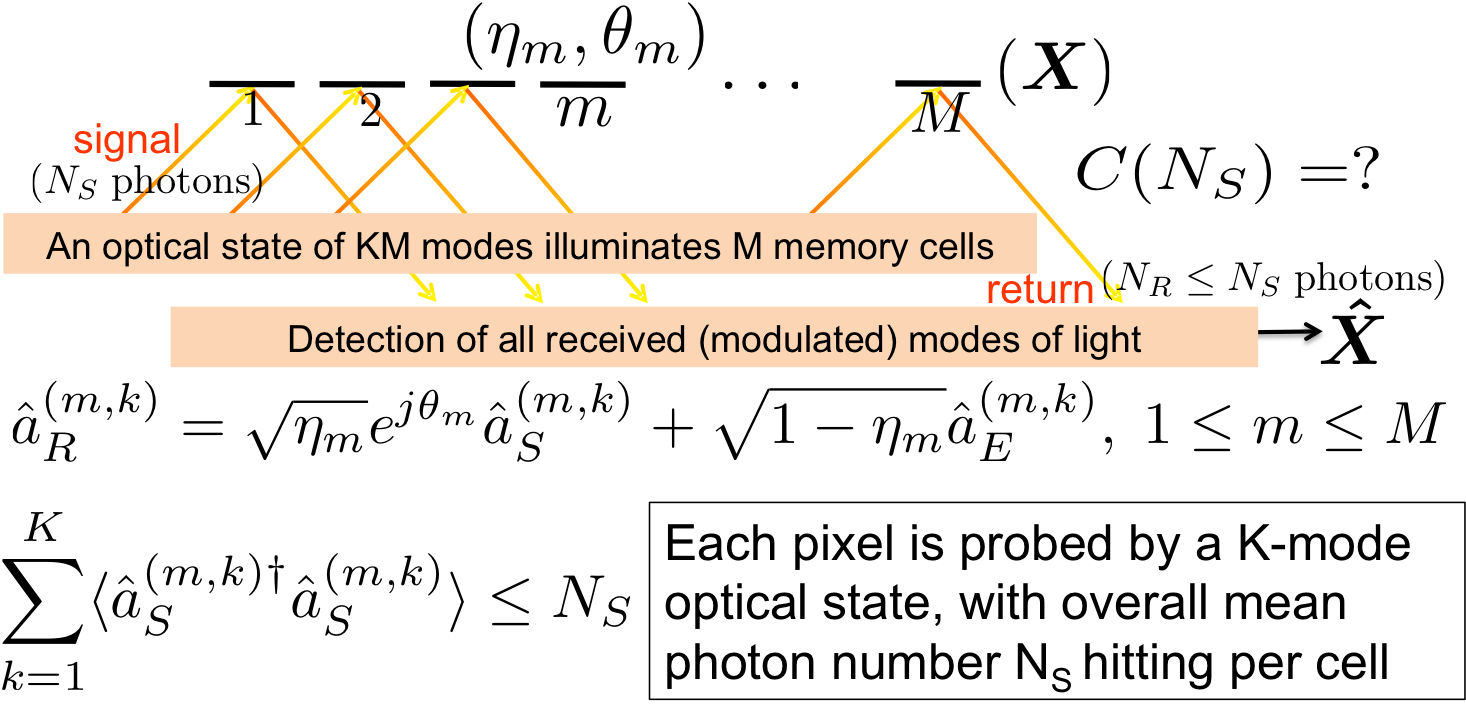}
\caption{Schematic of an optical memory that uses passive linear-optic reflective encoding. $M$ memory pixels are shown. Each pixel can modulate the spatial mode of the incident probe light by a power attenuation $\eta_m \in [0,1]$ and/or a (carrier) phase shift $\theta_m \in (0, 2\pi]$.}
\label{fig:setup}
\end{figure}

\section{Capacity of optical reading} \label{sec:capacity}

The setup we shall consider is shown in Fig.~\ref{fig:setup}. Each memory pixel is a reflective surface that can modulate the incident optical mode(s) by a power attenuation factor $\eta_m \in [0,1]$ and/or a (carrier) phase shift $\theta_m \in (0, 2\pi]$. A $K$-mode transmitter interrogates each memory pixel. Each pixel acts like a beam splitter, such that the $k$th return mode from the $m$th pixel is given by ${\hat a}_R^{(m,k)} = \sqrt{\eta_m}e^{j\theta_m}{\hat a}_S^{(m,k)}+\sqrt{1-\eta_m}{\hat a}_E^{(m,k)}$, where the $\{\hat{a}_S^{(m,k)}\}$ are the transmitter (signal) modes and the environment modes, $\{{\hat a}_E^{(m,k)}\}$, are taken to be in their respective vacuum states, implying no excess noise.  We will impose a mean photon-number constraint on the transmitter, $\sum_{k=1}^K\langle{\hat a}_S^{{(m,k)}\dagger}{\hat a}_S^{(m,k)}\rangle \le N_S$ photons per pixel. In what follows we will address the following two canonical questions:
\begin{enumerate}
\item {\bf Capacity}---How many bits of information can be reliably encoded and read per memory pixel, $C(N_S)$ bits/pixel, as a function of the average photon number spent to interrogate a pixel, $N_S$, when there are no constraints on the length of the code, the transmitter state and the receiver measurement? The photon information efficiency (PIE) is the number of bits read per signal photon, given by $C(N_S)/N_S$ bits/photon. As is true for most capacities, reading data at a rate $R < C(N_S)$ bits/pixel at a probability of word error $P_e^{(M)} \to 0$, may require coding over $M \to \infty$ many pixels and employing a JDR over infinitely many pixels.
 
\item {\bf Error exponent}---What is the minimum number of pixels $M$ required (length of code and JDR) to attain a certain PIE, such that $P_e^{(M)} \le \epsilon$?
\end{enumerate}

For both of the preceding performance metrics, we would also like to know by how much can non-classical states of light and/or non-standard optical receivers (including JDRs) outperform a coherent-state probe and the standard optical receivers (homodyne, heterodyne and direct-detection).

At this point, readers who are familiar with the Holevo capacity of bosonic channels~\cite{Gio04}, might see the correspondence between the above problem and the problem of finding the capacity $C_{\rm comm}(N_S)$ (bits/use) of a single-mode pure-loss bosonic channel when $N_S$ photons are transmitted, on average, per channel use. It is well known that the Holevo capacity for that channel---$C_{\rm comm}(N_S) = g(\eta N_S)$ bits/use, where $g(x) = (1+x)\log_2(1+x) - x\log_2(x)$ and $\eta$ is the channel's transmissivity---can be achieved using coherent-state modulation and a joint-detection receiver. It is easy to see that the capacity of optical reading, $C(N_S) \le C_{\rm comm}(N_S)$, because the light reflected from $M$ memory pixels can be regarded as an $M$-mode codeword.  It is \em not\/\rm\ obvious, however, that equality holds, i.e., $C(N_S) = C_{\rm comm}(N_S)$, nor is it clear that a coherent-state probe can attain the reading capacity.  It turns out that $C(N_S) = g(\eta N_S)$, where $\eta$ is now the average reflectivity of the encoded pixels, is \em only\/\rm\ possible for lossless optical reading, viz., phase-only encoding with $\eta=1$.  Furthermore, even in the lossless case, $C(N_S) = g(N_S)$ \em cannot\/\rm\ be achieved with coherent states~\cite{Guh11, Guh12a}. Thus, despite the similarities of optical communication and optical reading, the latter problem is more constrained than the former, because its modulation and coding happen passively at the pixels, with the transmitter being ignorant of the information to be modulated on the probe light. In communication---for which the information to be modulated and coded is available to the transmitter---the spate of recent work on Holevo capacity-achieving codes~\cite{Wil12, Wil12a} and joint-detection receivers~\cite{Guh11a, Llo10, Wil12b, Chu11} has yet to yield an efficient Holevo-capacity-achieving code and a structured optical design for its JDR.  Surprisingly, we will exhibit an explicit capacity-achieving system for lossless optical reading.    

\begin{figure}
\centering
\includegraphics[width=1.1\columnwidth]{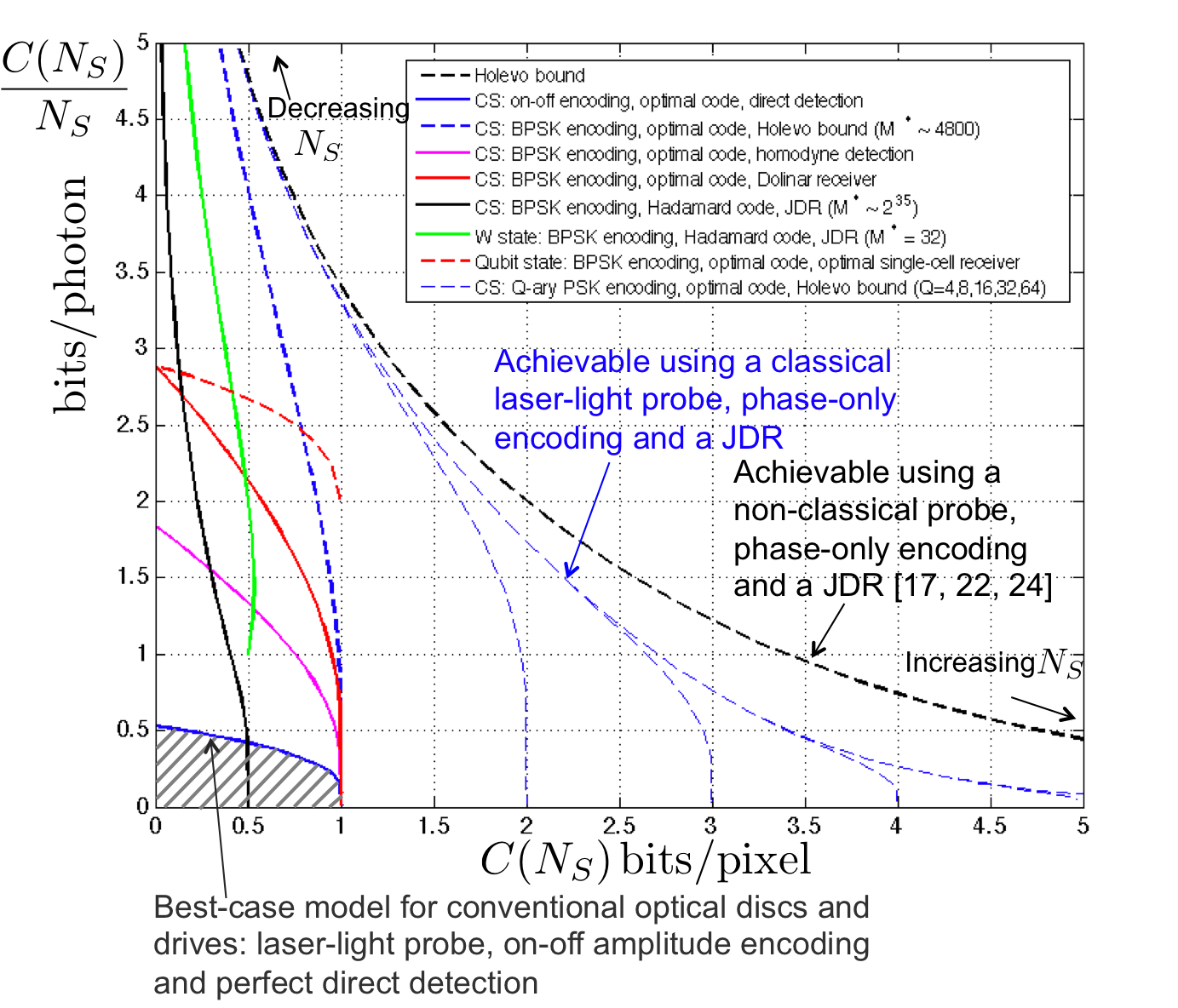}
\caption{Photon information efficiency of optical reading (bits read per photon) versus the encoding efficiency (bits encoded per pixel) for various transmitter and receiver strategies. The gray shaded area shows the Shannon limit of the performance achievable using an ideal laser-light probe, on-off amplitude encoding and an ideal direct detection receiver---an overly-optimistic model for how conventional optical drives read data from disks.
}
\label{fig:PIE_vs_SE}
\end{figure}

To focus on the fundamental aspects of the capacity and error-exponent questions, we will assume that there is: no return-path loss of the probe light (except for any loss due to amplitude modulation by the memory pixels); no excess noise (such as noise due to detector imperfections or a thermal background); and a diffraction-limited transceiver with spatially-resolved pixels. There is a fundamental trade-off between optical reading's photon efficiency (bits read per photon) and its encoding efficiency (bits encoded per pixel). This trade-off---for a variety of transmitter, encoding, and receiver techniques---is summarized in Fig.~\ref{fig:PIE_vs_SE}. High photon efficiency (hence, low encoding efficiency) is attained for a low-brightness transmitter (low $N_S$), whereas attaining high encoding efficiency (hence, low photon efficiency) requires a bright transmitter (high $N_S$). The gray-shaded area in Fig.~\ref{fig:PIE_vs_SE} corresponds to the best performance achievable using conventional techniques. The derivations of these trade-off curves will be explained below. We will focus on the high photon efficiency (low photon flux) regime in this paper. It is here that the advantage of joint detection is the most pronounced. We will defer consideration of the high encoding efficiency regime to the sequel~\cite{Guh12a} of the present work; it will be a long version of~\cite{Guh11}. Finally, we will also limit our scope in this paper to a single-mode transmitter, i.e., $K=1$. 

\begin{figure}
\centering
\includegraphics[width=\columnwidth]{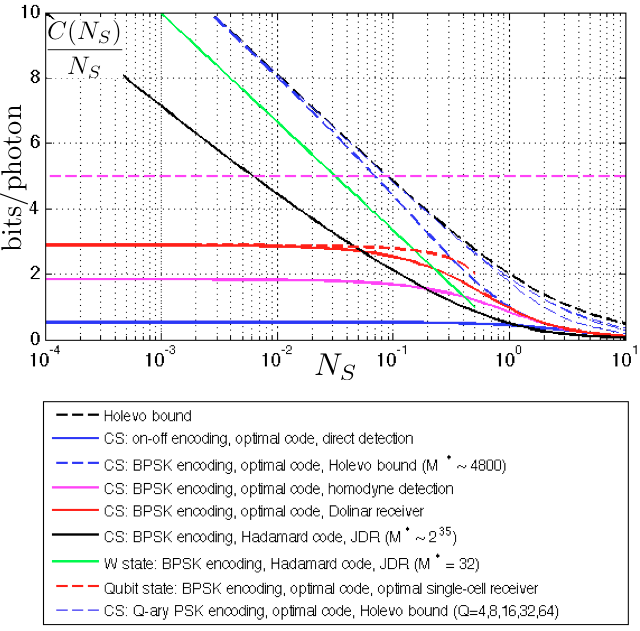}
\caption{Photon information efficiency (PIE) versus the mean photon number $N_S$ used to interrogate each memory pixel. CS denotes coherent state. $M^* =$ Minimum number of pixels needed to achieve $5$ bpp at $P_e = 10^{-3}$.
}
\label{fig:PIE}
\end{figure}

We begin by considering an idealized model for the standard CD/DVD drive, i.e., a laser-light probe, on-off amplitude modulation, and a direct-detection receiver.  Interrogation and detection of each pixel induces a binary asymmetric channel, as shown in Fig.~\ref{fig:CDmodel_bc}(a). The Shannon capacity~\cite{Sha48} of this channel is given by:
\begin{figure}
\centering
\includegraphics[width=\columnwidth]{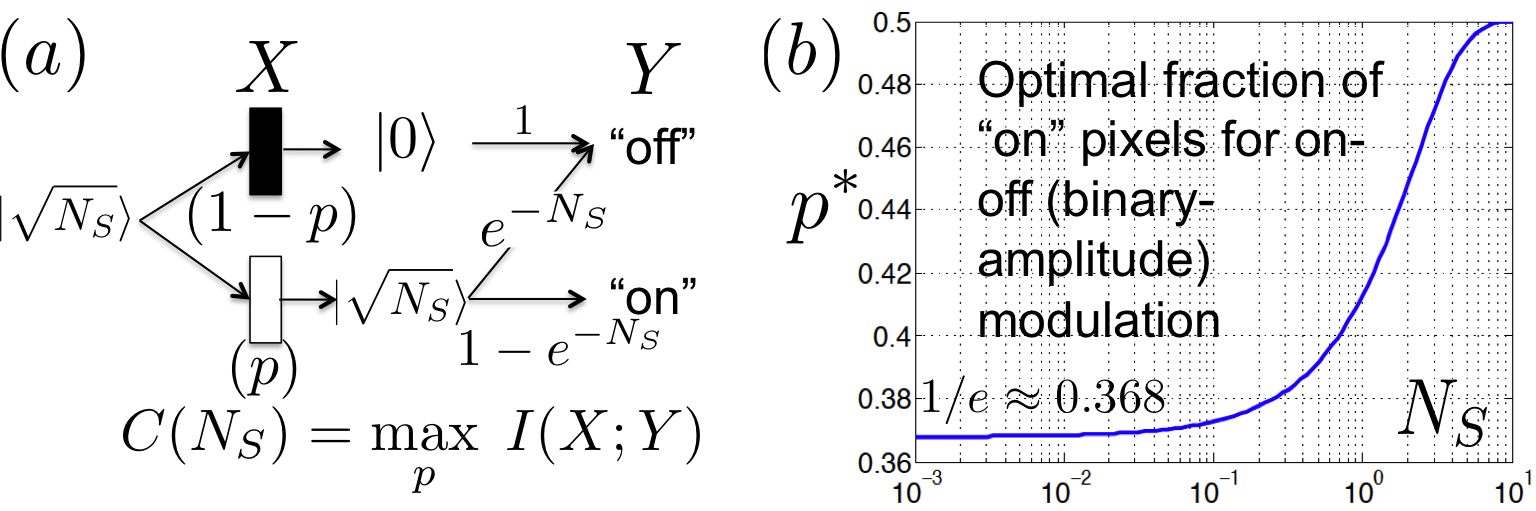}
\caption{(a) The induced binary channel for a coherent-state probe, on-off pixel encoding, and shot-noise-limited direct detection. (b) The optimal fraction of ``on" pixels $p^*$ that maximizes the number of bits read per pixel $C(N_S)$, when mean photon number $N_S$ is used to interrogate each memory pixel.}
\label{fig:CDmodel_bc}
\end{figure}
\begin{eqnarray}
C(N_S) &=& \max_{p \in (0,1)}I(X;Y)\\
&=& \max_{p \in (0,1)}\left[H(Y) - H(Y|X)\right]\\
&=& \max_{p \in (0,1)}\left[H\!\left(p(1-e^{-N_S})\right)-pH\!\left(e^{-N_S}\right)\right],\label{eq:OOKcapacity}
\end{eqnarray}
where $H(x) = -x\log_2(x) - (1-x)\log_2(1-x)$ is the binary entropy function. The optimal value of $p$ that maximizes the mutual information $I(X;Y)$ is the fraction of ``on" pixels in a capacity-achieving code, which is readily computed to be:
\begin{equation}
p^*(N_S) = \frac{1}{\left(1-e^{-N_S}\right)\left[1+2^{H\left(e^{-N_S}\right)/\left(1-e^{-N_S}\right)}\right]}.
\end{equation}
Figure~\ref{fig:CDmodel_bc}(b) shows that $p^*(N_S) \to 0.5$ for $N_S \gg 1$. This is the regime in which a standard CD drive operates, wherein optimal codes have equal fractions of {\em on} and {\emÊoff} pixels. On the other hand, $p^*(N_S) \to 1/e \approx 0.368$ for $N_S \ll 1$. The solid blue line in Fig.~\ref{fig:PIE} plots the PIE, $C(N_S)/N_S$, as a function of $N_S$ for on-off pixel modulation, a coherent-state probe, and direct detection. The PIE caps off, $C(N_S)/N_S \to 1/e \ln (2) \approx 0.53$ bits per photon (bpp) for $N_S \ll 1$. Thus, even with the optimal code (codewords infinitely many pixels long), using on-off modulation, an ideal laser transmitter, and an ideal direct-detection receiver, no more than about 0.5\,bits can be read per transmitted photon.

\begin{figure}
\centering
\includegraphics[width=0.8\columnwidth]{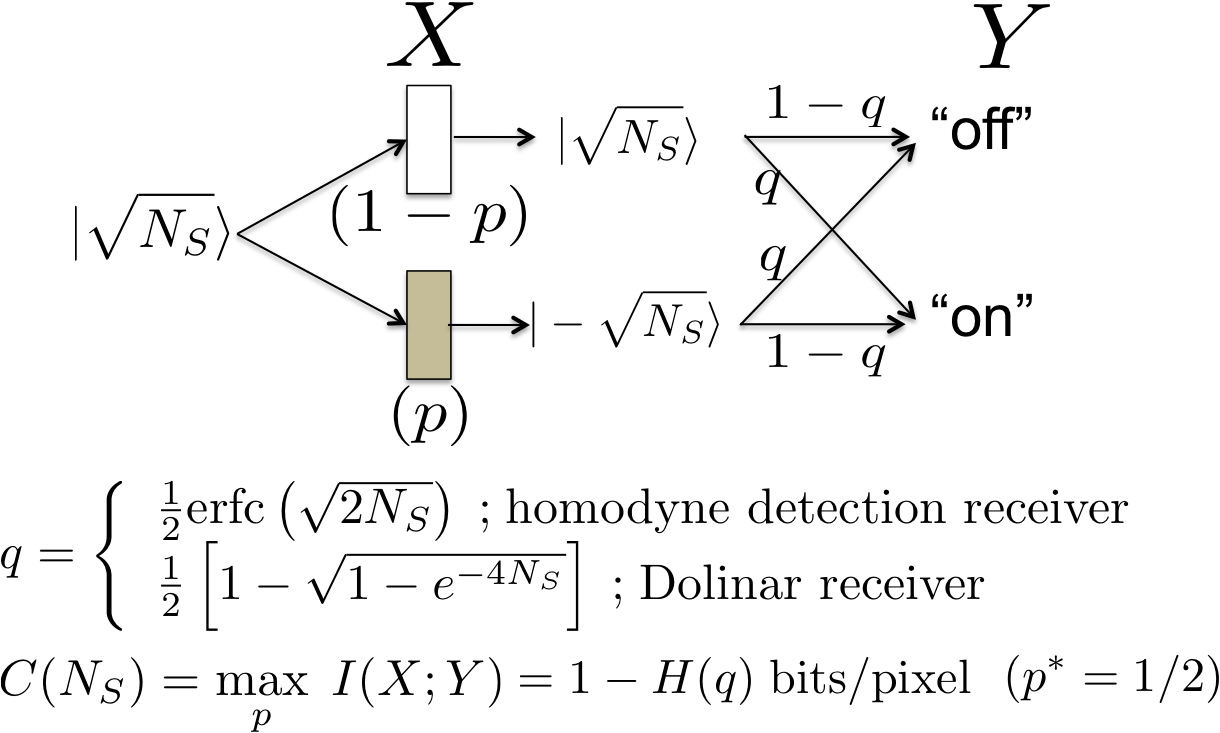}
\caption{A binary symmetric channel is induced when a coherent state probe $|\sqrt{N_S}\rangle$ interrogates each memory pixel, and the reflected light is detected by either a conventional homodyne receiver with a threshold detector, or by the Dolinar receiver---a receiver that can attain the minimum probability of error for discriminating between two coherent states.}
\label{fig:BSC}
\end{figure}

Let us now consider the binary phase-shift keyed (BPSK) modulation. Each memory pixel is a perfectly reflective pixel but some are etched $\lambda/2$ deeper into the surface of the disc, where $\lambda$ is the center-wavelength of the (quasimonochromatic) probe light. A coherent-state probe $|\sqrt{N_S}\rangle$ sent to interrogate the $m${\rm th} pixel gets reflected as $|\sqrt{N_S}\rangle$ or $|-\sqrt{N_S}\rangle$ depending upon whether that pixel's phase is $\theta_m = 0$ or $\pi$. The conventional receiver to discriminate the states $\left\{|\sqrt{N_S}\rangle, |-
\sqrt{N_S}\rangle\right\}$ uses homodyne detection, which results in a Gaussian-distributed measurement outcome $\beta$, with mean $\pm \sqrt{N_S}$ and variance 1/4.  The minimum error-probability post-detection processor is the threshold test. $\beta \ge 0 \Rightarrow \theta = 0$ and $\beta < 0 \Rightarrow \theta = \pi$, which induces a binary symmetric channel (BSC) with crossover probability $q_{\rm hom} = {\rm erfc}(\sqrt{2N_S})/2$ (see Fig.~\ref{fig:BSC}), whose capacity is given by $C(N_S) = 1-H(q_{\rm hom})$ bits/pixel, and is achieved for an equal prior ($p^* = 1/2$) for the two phase values. The minimum achievable error-probability for discriminating a single copy of the two equally-likely coherent states $\left\{|\sqrt{N_S}\rangle, |-\sqrt{N_S}\rangle\right\}$ is given by the Helstrom limit~\cite{Hel76}, $P_{e,\min} = \left[1-\sqrt{1-e^{-4N_S}}\right]/2$. This minimum probability of error can in principle be achieved exactly using the Dolinar receiver~\cite{Dol76, Coo07}, which is a structured optical receiver design that uses a local time-varying optical feedback and high-speed ideal single-photon detection. The Dolinar receiver used with BPSK modulation induces a BSC with crossover probability $q_{\min} = P_{e,\min}$, and capacity $C(N_S) = 1-H(q_{\min})$ bits/pixel. The magenta and red plots in Fig.~\ref{fig:PIE} show the PIE for BPSK encoding with the homodyne and Dolinar receivers, which cap off at $4/\pi\ln (2) \approx 1.84$\,bpp and $2/\ln (2) \approx 2.89$\,bpp, respectively, for $N_S \ll 1$.

For a single $\left\{0,\pi\right\}$ binary phase-modulated pixel, of all (multimode) transmitter states with mean photon number $N_S$, the one that minimizes the probability of error is the single-mode ($K=1$) {\em single-rail encoded} qubit state, $|\psi\rangle^S = \sqrt{1-N_S}|0\rangle + \sqrt{N_S}|1\rangle$, which attains zero probability of error for $N_S \ge 1/2$, and $P_{e,\min,{\rm QS}} = \left[1-\sqrt{1-[1-2N_S]^2}\right]/2$, for $N_S < 1/2$~\cite{Nai12}. Capacity is again given by the BSC capacity formula, $C(N_S) = 1-H(P_{e,\min,{\rm QS}})$ bits/pixel (see the red-dashed plot in Fig.~\ref{fig:PIE}). The resulting PIE caps off at $2/\ln (2) = 2.89$\,bpp for $N_S \ll 1$. Note that the BPSK pixel modulation format achieves the minimum possible probability of error over all transmitters and receivers acting on reflection from single pixels, and hence achieves higher capacity (and PIE) than what can be obtained via amplitude modulation alone~\cite{Pir11a}. Notwithstanding, the PIEs of all the cases considered above cap off below $3$\,bpp.

The classical information-carrying (Holevo) capacity of a quantum signaling alphabet was found by Holevo, Schumacher and Westmoreland~\cite{Hol79,Hau96}. The Holevo capacity of the pure-loss (vacuum-noise) optical channel with a mean received photon number per mode $N_S$ is $g(N_S)$ bits/mode~\cite{Gio04}. This capacity is achievable using a coherent-state code with symbols $|\alpha\rangle$ chosen in an independent, identically-distributed (i.i.d.) manner from the isotropic Gaussian distribution, $p(\alpha)=e^{-|\alpha|^2/N_S}/{\pi N_S}$. Hence, for communicating classical data on a pure-loss optical channel, non-classical transmitter states cannot achieve any higher capacity than coherent states. From the capacity theorem converse in~\cite{Gio04}---treating the reflected light from the memory pixels as a modulated codeword---and monotonicity of the $g(\cdot)$ function, the capacity of optical reading must satisfy the upper bound, $C(N_S) \le g(N_S)$, for all single-mode probe states ($K=1$). However, the reading problem has less encoding freedom than the communication transmitter, because its modulation must be passive (non-amplifying) at the pixels. That is why $C(N_S)=g(N_S)$ bits/pixel is not achievable for optical reading using a coherent-state transmitter~\cite{Guh12a}. However, we have shown that $C(N_S)=g(N_S)$ bits/pixel {\em is} achievable using a non-classical transmitter~\cite{Guh11, Guh12a} and a sequential-decoding quantum joint-detection receiver~\cite{Wil12b}.

The black-dashed line in Fig.~\ref{fig:PIE} shows the PIE of the Holevo bound $g(N_S)/N_S$. Note that, unlike all the capacity results for on-off and binary-phase modulation with explicit pixel-by-pixel detection schemes considered above, the Holevo bound has no upper limit to the number of bits that can be read per expended photon. However, the higher the desired PIE, the lower must be the mean photon number $N_S$ used to interrogate each pixel, resulting in a lower data rate $C(N_S)$ (bits/pixel) read. Even though coherent states do not achieve the Holevo bound on reading capacity, a coherent-state probe can nevertheless approach $g(N_S)$ in the high-PIE ($N_S \ll 1$) regime when employed in conjunction phase modulation and the optimal JDR that makes a collective measurement over return modes from many pixels,  as we will now show. The blue-dashed plots in Fig.~\ref{fig:PIE} are the Holevo-limit PIEs of $Q$-ary phase-shift-keying (PSK) constellations used to encode the data for $Q=2,4,8,16,32$.  Because the PSK Holevo limit for any $Q$ is an achievable rate~\cite{Hau96}, it is a lower bound to the reading capacity, i.e., $C(N_S) \ge C_{\rm PSK-Holevo}(N_S)$, where
\begin{equation}
C_{\rm PSK-Holevo}(N_S) = \max_{Q \ge 2}\;-\sum_{q=1}^Qy_q(N_S)\log_2y_q(N_S),\nonumber
\end{equation}
with $\left\{y_q(N_S)\right\}$, $1 \le q \le Q$, being the probability distribution,
\begin{eqnarray}
\lefteqn{y_q(N_S) = } \nonumber \\[.05in]
&&\frac{1}{Q}\sum_{k=1}^Qe^{-{N_S}\left(1-\cos\left[\frac{2\pi{k}}{Q}\right]\right)}\cos{\left[{N_S}\sin\left[\frac{2\pi{k}}{Q}\right]-\frac{2\pi{k}q}{Q}\right]}.\nonumber
\end{eqnarray}
For $Q=2$ (BPSK modulation), the Holevo capacity is given by $C_{\rm BPSK}(N_S) = H\left((1+e^{-2N_S})/2\right)$ bits/pixel.   Its PIE is shown by the dark dashed-blue plot in Fig.~\ref{fig:PIE}, where it is seen to approach the Holevo limit $g(N_S)/N_S$ at low $N_S$. Thus, the gap between the PIE of BPSK modulation used with an optimal single-symbol receiver (solid-red plot) and the Holevo limit of BPSK (dashed-blue plot) can and must be bridged using a JDR.

\begin{figure}
\centering
\includegraphics[width=\columnwidth]{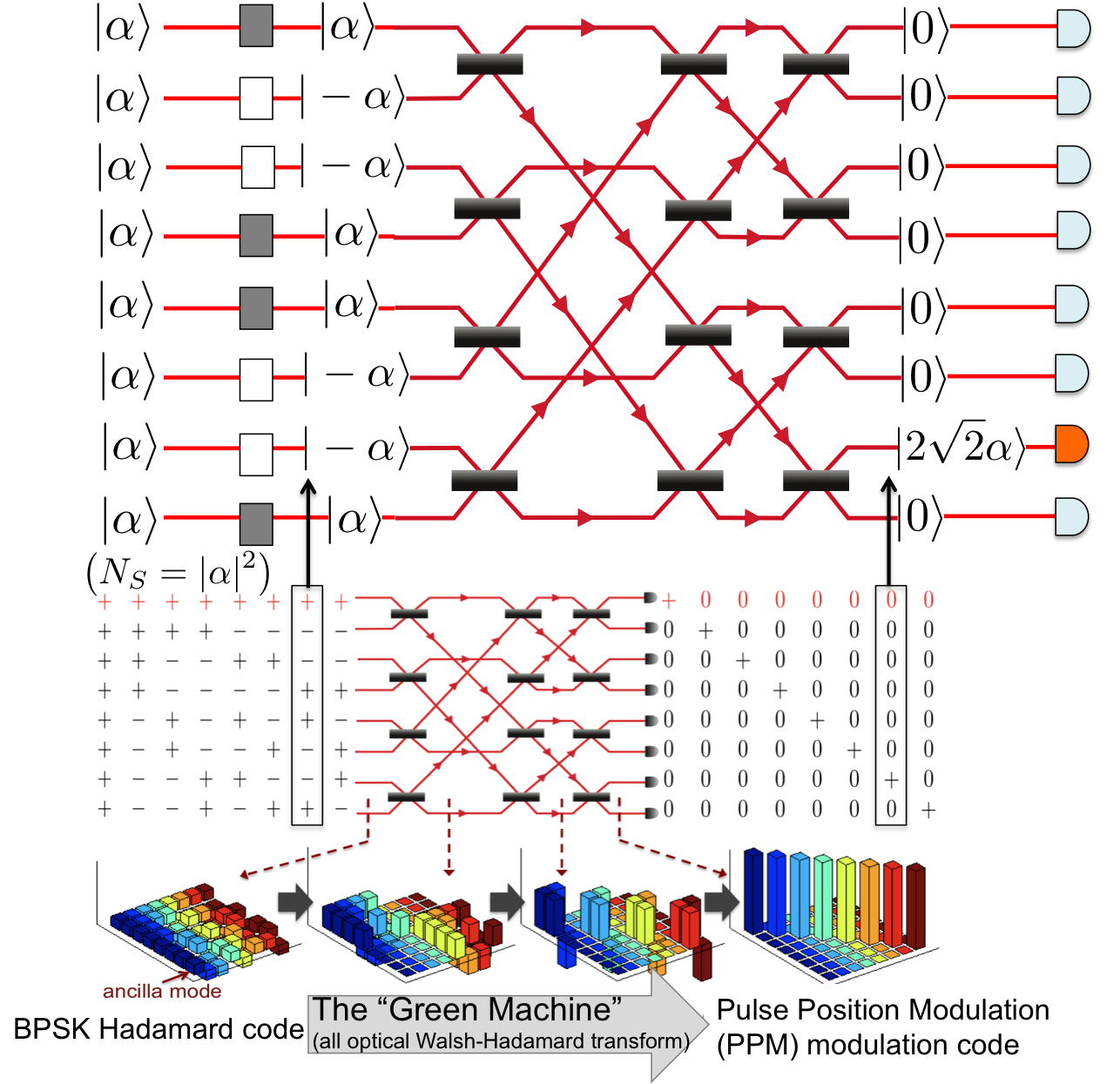}
\caption{The Green Machine JDR. Each vertical column of `+'s and `-'s is a reflection from $M$ binary-phase-coded memory pixels---a coherent-state BPSK codeword from the Hadamard code with $M$ pulses of mean photon number $N_S$ each. Each of the $M$ codewords transforms into exactly one coherent state of mean photon number $MN_S$ at one distinct output port of the optical circuit of $(M/2)\log_2(M)$ beam splitters. Under ideal conditions, a click at one of the $M$ single-photon detectors identifies the reflected codeword with no error, whereas a no-click leads to an {\em erasure} outcome, which induces the $M$-ary symmetric erasure channel shown in Fig.~\ref{fig:GMchannelmodel}.}
\label{fig:GM}
\end{figure}

\begin{figure}
\centering
\includegraphics[width=\columnwidth]{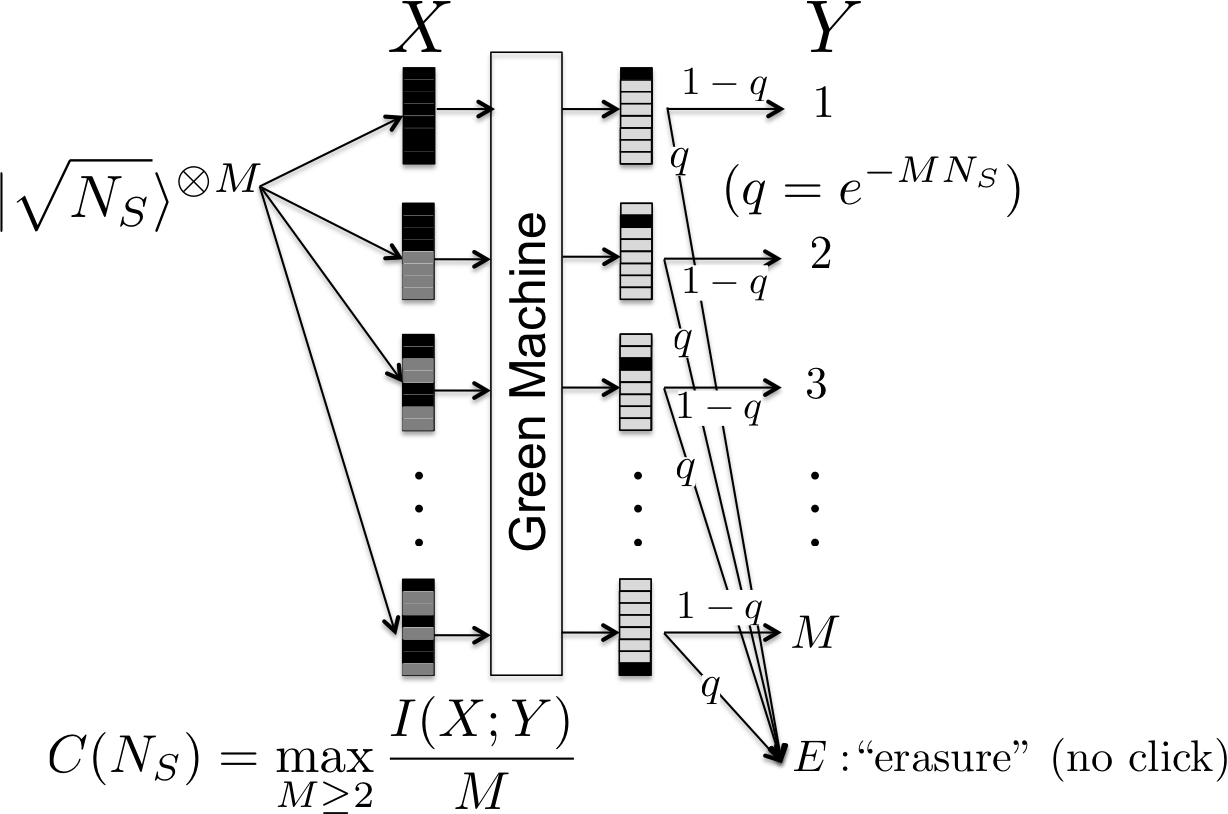}
\caption{The $M$-input, $M+1$-output channel induced by the coherent-state probe, binary-phase Hadamard coded memory, and the Green Machine JDR.}
\label{fig:GMchannelmodel}
\end{figure}

The first explicit code-JDR pair for a BPSK alphabet that achieves {\em superadditive} capacity (i.e., higher capacity than what is achievable with the optimal single-symbol receiver for BPSK) was found by one of us recently in the context of communication~\cite{Guh11a}, but that construct---which we will now describe---also applies to optical reading. The Green Machine JDR for BPSK-modulated pixels and a coherent-state probe is depicted in Fig.~\ref{fig:GM}. It uses a $(2^m,2^m,2^{m-1})$ binary Hadamard code to encode the binary phases on $M = 2^m$ pixels. The receiver comprises an interferometer made of $(M/2)\log_2(M)$ 50-50 beam splitters arranged in a format---first envisioned by R. R. Green as a classical decoding circuit for Hadamard codes~\cite{Gre66}---that interferometrically mixes the modulated light from the $M$ pixels, transforming the BPSK Hadamard codeword, through $\log_2(M)$ stages of the Green Machine, into a spatial pulse-position-modulation (PPM) code. A coherent-state pulse with mean photon number $MN_S$ appears at one of the $M$ outputs, depending upon which of the $M$-pixel Hadamard codewords the probe light interrogates. The output is detected by an array of $M$ signal-shot-noise-limited single-photon detectors. This probe-code-JDR combination induces an $M$-input, $M+1$-output superchannel, shown in Fig.~\ref{fig:GMchannelmodel}, whose capacity (in bits/pixel) is given by:
\begin{eqnarray}
&&C_{\rm BPSK-Hadamard-JDR}(N_S) =\max_{M \ge 2}\frac{I(X;Y)}{M}\label{eq:GMcapacity}\\
&=&\max_{M \ge 2}\frac{\log_2(M)\left(1-e^{-MN_S}\right)}{M} \nonumber \\
&=& \frac{1}{\ln (2)}\left[N_S\ln\!\left(\frac{1}{N_S}\right) - N_S\ln\!\left[\ln\!\left(\frac{1}{N_S}\right)\right] + \ldots \right], \nonumber
\end{eqnarray}
when $N_S \ll 1$. Here, the PIE-maximizing value of the code size $M$ as a function of $N_S$ is given by $M^* \approx -5/2N_S\ln (N_S)$, for $N_S \ll 1$. This PIE is plotted as the solid-black line in Fig.~\ref{fig:PIE}. Unlike all the structured probe-receiver cases we have considered so far---in which the optical receiver measured the reflected light from each pixel individually---the PIE attained by the BPSK Hadamard code and the Green Machine JDR increases without bound as $N_S \to 0$. Note that this PIE is optimal to the leading-order term of the Holevo bound (both the unrestricted-modulation Holevo bound and the coherent-state-probe BPSK-encoding Holevo capacity) for $N_S \ll 1$:
\begin{equation}
C(N_S) = \frac{1}{\ln (2)}\left[N_S\ln\!\left(\frac{1}{N_S}\right) + N_S + \ldots \right] \;{\text{bits/pixel}}. 
\end{equation}
One can increase the photon efficiency slightly by using the $(2^m-1,2^m,2^{m-1})$ Hadamard code, thereby using one less ($M = 2^m-1$) pixel, and retaining a local-oscillator reference at the transmitter for use as a local input into the Green Machine. Note that the achievable capacity in Eq.~\eqref{eq:GMcapacity} and all the coherent-state structured-receiver capacities given above are Shannon capacities of the respective discrete memoryless channels induced by the choice of the probe-code-receiver combination. Hence, in order to achieve error-free reading at a rate close to these capacities (in bits/pixel), a suitable Shannon-capacity-approaching outer code---such as a Reed Solomon code---will be required.


\begin{figure}
\centering
\includegraphics[width=\columnwidth]{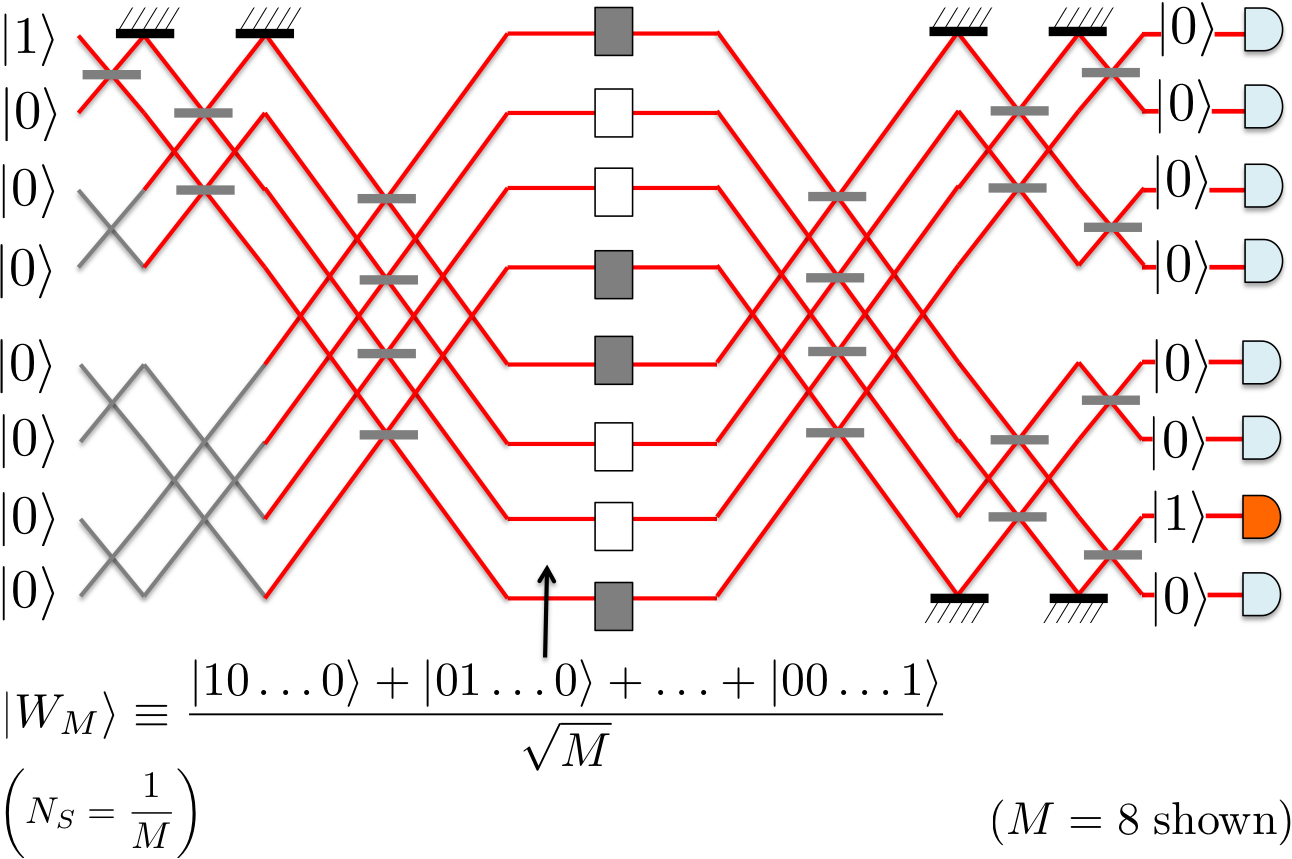}
\caption{The W-state transmitter, generating the $M=8$ mode W state, interrogates $M$ binary-Hadamard phase-coded pixels. The wave function of the single photon evolves through the receiver stages, eventually coalescing into the single-photon Fock state $|1\rangle$ at one of the $M$ outputs of the receiver, depending upon which one of the $M$ Hadamard codewords the transmitter state interrogated.}
\label{fig:Wstate_transmitter}
\end{figure}
\begin{figure}
\centering
\includegraphics[width=\columnwidth]{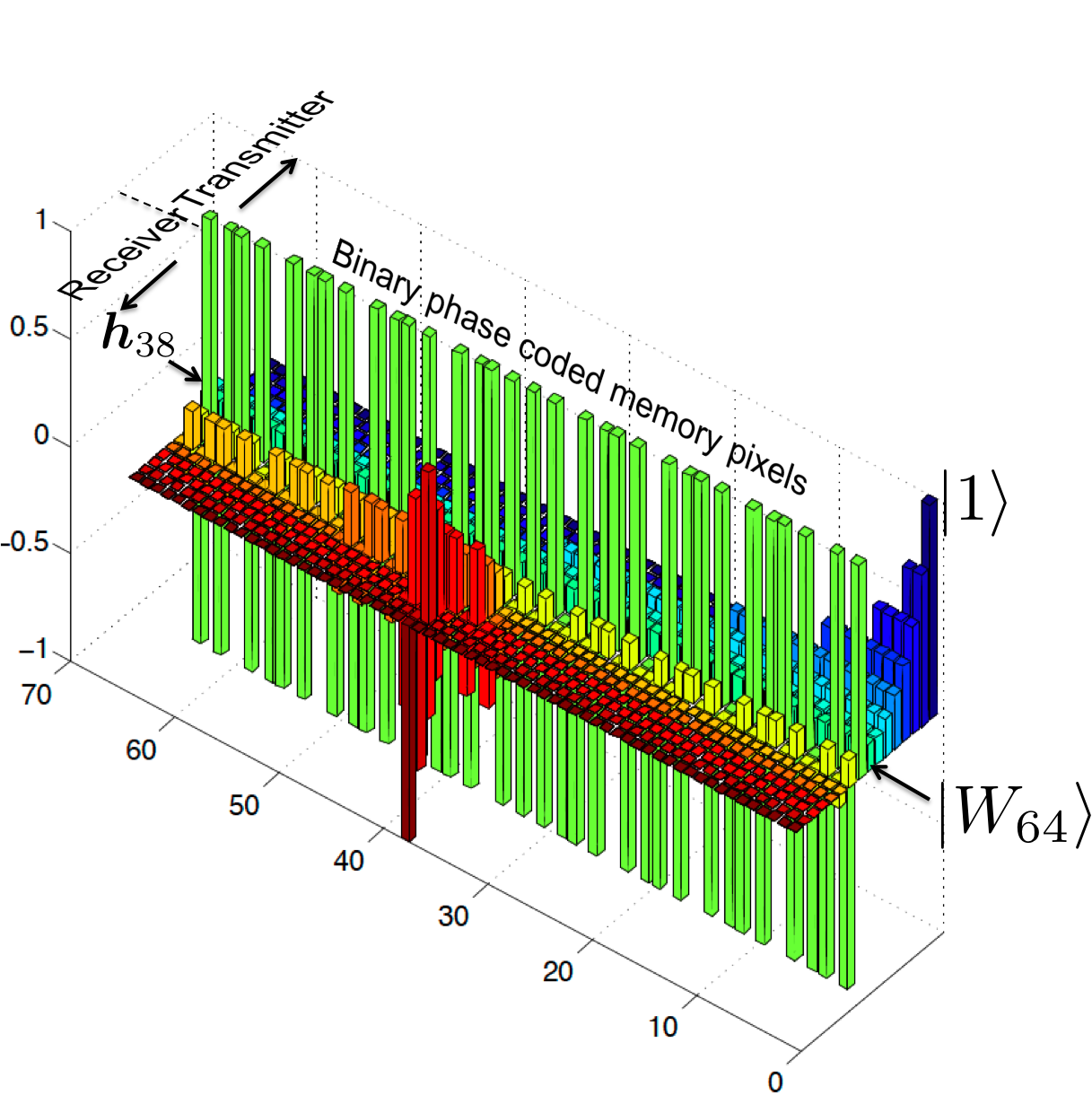}
\caption{An $M=64$ example of W-state probability-amplitude evolution during the encoding and decoding phases of optical reading. The vertical axis denotes the amplitude and binary phase of the photon wave-function. Positive (upwards from $0$) denotes $0$-phase and negative (downwards from $0$) denotes $\pi$-phase. The green vertical bars depict the $38{\rm th}$ of the BPSK Hadamard codewords, ${\boldsymbol h}_m$, $1 \le m \le 64$, using the above sign-convention for $0$ and $\pi$ phases. A single-photon Fock state $|1\rangle$ is shown to go through the $\log_2(M) = 8$ stages of the encoding circuit, shown in Fig.~\ref{fig:Wstate_transmitter}, to form the equal-superposition W-state $|W_{64}\rangle$, which undergoes phase-modulation at the memory pixels (green bars). The modulated W-state is shown to go through the $8$ stages of the optical receiver circuit, also shown in Fig.~\ref{fig:Wstate_transmitter}, eventually forming a single-photon Fock state $|1\rangle$ (with an unimportant phase) at the correct ($38{\rm th}$) output port, which is then detected error-free by an ideal single-photon detector.}
\label{fig:Wstate_14_38}
\end{figure}

Now, let us keep the same BPSK modulation and Hadamard code but consider using a spatially-entangled non-classical probe state, the W-state, instead of a coherent state. This probe sends exactly one photon in a coherent superposition of $M$ spatial modes,
\begin{equation}
|W_M\rangle \equiv \frac{|10\ldots 0\rangle + |01\ldots 0\rangle + \ldots + |00\ldots 1\rangle}{\sqrt{M}},
\end{equation}
to interrogate $M$ memory pixels, where $M$ is taken to be even.  It can be prepared using a single-photon source (generating a one-photon Fock state $|1\rangle$) split via an array of 50-50 beam splitters as shown in Fig.~\ref{fig:Wstate_transmitter}. Recently, it was shown how to perform fast heralded generation of the W-state, and other complicated mode-shaped single photon states, by indirectly tailoring the mode of the single photon via amplitude modulation of the classical pump field driving a spontaneous parametric downconversion process~\cite{Kop11}.  Reflection of the W-state by the Hadamard-phase-coded pixels causes the `$+$' signs in the W-state superposition corresponding to the pixels with $\theta = \pi$ to flip to `$-$' signs. Let the memory-modulated state for codeword $m$ be $|W_M^{(m)}\rangle$, for $1 \le m \le M$. Because any pair of codewords from the Hadamard code differ in exactly $M/2$ positions, the $\{|W_M^{(m)}\rangle\}$ are mutually orthogonal quantum states; i.e., $\langle W_M^{(m_1)}|W_M^{(m_2)}\rangle = \delta_{m_1,m_2}$. Therefore, it is possible, in principle, to discriminate these $M$ modulated states with zero probability of error. An explicit receiver that accomplishes this zero-error discrimination is shown in Fig.~\ref{fig:Wstate_transmitter}. The wave function of the single photon evolves through the $\log_2(M)$ receiver stages, eventually coalescing into the single-photon Fock state $|1\rangle$ at one of the $M$ outputs of the beam-splitter circuit depending upon which one of the $M$ Hadamard codewords the transmitter state interrogated. A single-photon Fock state $|1\rangle$ generates a click with probability $1$ when detected by an ideal single-photon detector (unlike a coherent state $|\beta\rangle$, which generates a click with probability $1 - e^{-|\beta|^2}$ under ideal conditions). Therefore, the W-state transceiver reads $\log_2(M)$ bits of information {\em error-free}---i.e., without any further outer coding---using just one transmitted photon, with no upper limit on $M$. Clearly, $N_S = 1/M$. Therefore, capacity is given by:
\begin{equation}
C_{\rm W-state}(N_S) = N_S\log_2\!\left(\frac{1}{N_S}\right)\;{\text{bits/pixel}}.
\end{equation}
This capacity has the same low-$N_S$ leading-order term as the Holevo limit and the coherent-state Green Machine JDR. The PIE for the W-state is exactly $\log_2(1/N_S)$ bpp, i.e., a straight line when plotted versus $N_S$ on a logarithmic scale (see the green line in Fig.~\ref{fig:PIE}). Figure~\ref{fig:Wstate_14_38} shows an $M=64$ example of the single photon's probability-amplitude evolution during the encoding and decoding phases of optical reading using a W-state. 

\section{Error exponent of quantum reading}\label{sec:error exponents}

All the results obtained in Section~\ref{sec:capacity} are Holevo or Shannon capacities. Thus, achieving a reliable rate of reading---i.e., reading information such that the probability of codeword error $P_e^{(M)} \le \epsilon$ for some low-enough threshold $\epsilon$---at any given rate $R < C(N_S)$ bits/pixel would require an optimal outer code for all the Shannon-capacity/structured-receiver cases considered, and would require an optimal code as well as an optimal JDR for the Holevo-capacity results. The W-state example does not require an outer code, because of its zero-error receiver.

Reading capacity gives a crucial information-theoretic perspective, namely, the fundamental limit on achievable rates at which data can be read. However, capacity alone only specifies the maximum achievable rate.  It provides no information about the coding and receiver complexity required to read data reliably at any achievable rate. Hence, a stronger form of the channel coding theorem has been pursued to determine the behavior of the minimum codeword-error probability, $P_e^{(M)}$, as a function of the codeword length (number of pixels) $M$ and the data rate $R$ (bits/pixel), for all rates $R < C$---both for classical channels (where $C$ is the channel's Shannon capacity)~\cite{Sha59,Gal68} as well for quantum channels (where $C$ is the channel's Holevo capacity)~\cite{Bur98,Dal12}. We define the reliability function or the {\em error exponent} for optical reading as~\cite{Sha59},
\begin{equation}
E(R) \equiv \limsup_{M \to \infty}\frac{-\ln [P_e^{\rm opt}(R,M)]}{M}, \forall R < C(N_S),
\end{equation}
where $P_e^{\rm opt}(R,M)$ is the average word-error probability for the optimal block code of $M$ pixels and rate $R$. The error exponent describes how quickly the error probability decays as a function of $M$, and hence serves to indicate how difficult it may be to achieve a certain level of reliability in reading at a given rate below the capacity. Although it is difficult to exactly evaluate $E(R)$, its classical lower bound is available due to Gallager~\cite{Gal68}. This lower bound to the error exponent is known as the random-coding lower bound, and has been used to estimate the codeword length required to achieve a prescribed error probability for various communication settings. Burnashev and Holevo found the random-coding bound and the expurgated bound for sending classical data on quantum channels, both being lower bounds to $E(R)$ for a pure-state alphabet~\cite{Bur98}, and later generalized the expurgated bound to a mixed-state alphabet~\cite{Hol00}. The best known lower bound to the quantum channel's reliability function $E(R)$ was reported by Hayashi~\cite{Hay07}. For classical channels, there exists an upper bound (the sphere-packing bound) which coincides with $E(R)$ for high rates, i.e., for rates $R$ close to the Shannon capacity, $C$, and thus gives the exact expression for $E(R)$. Until very recently, no useful upper bound for $E(R)$ had been known for the quantum case. That changed, however, when Dalai reported the sphere-packing bound on the error exponent for sending classical data over a quantum channel~\cite{Dal12}. Dalai's upper bound to $E(R)$ for the quantum channel coincides with the random-coding lower bound at high rates, thereby yielding the true value of $E(R)$ in this region. More work needs to be done in the low-rate regime, in order to fully determine the error exponent $E(R)$ for a quantum channel for all rates $R$ below the Holevo capacity.

In order to compare the error-exponent performance of various transceivers we proposed in Section~\ref{sec:capacity}, let us choose a PIE goal of $5$\,bpp, and probability of word error threshold $\epsilon = 10^{-3}$. 

\subsection{Coherent-state probe: optimal code, optimal JDR}
\begin{figure}
\centering
\includegraphics[width=\columnwidth]{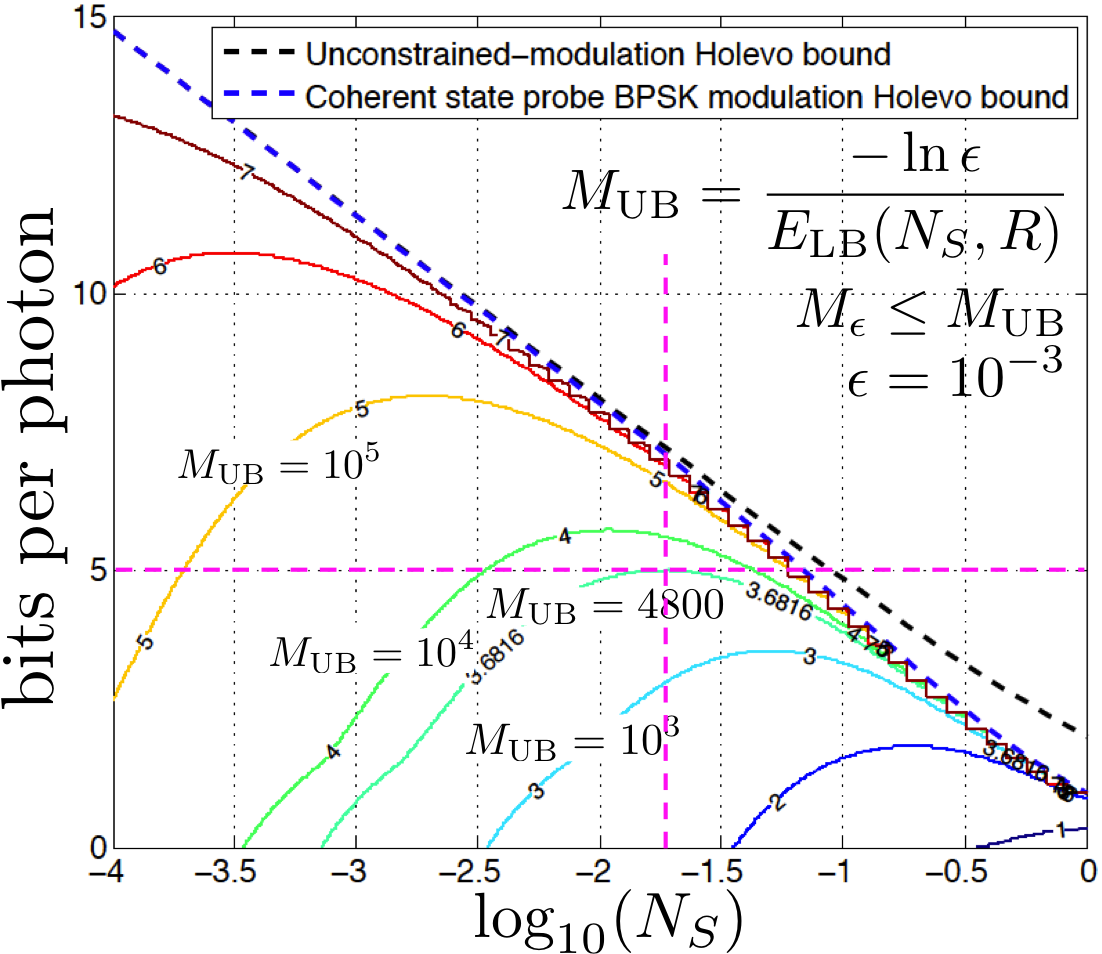}
\caption{Contours of constant $M_{\rm UB} \equiv -\ln(\epsilon)/E_{\rm LB}(N_S,R)$---the upper bound to the number of pixels required to achieve 5\,bpp with a $10^{-3}$ word-error probability found from the Burnashev-Holevo random-coding bound for a pure-state quantum channel---plotted in the PIE ($R/N_S$) vs. $N_S$ plane. A coherent-state probe interrogating a binary-phase coded memory and an optimal JDR are assumed.}
\label{fig:EE_CSBPSK}
\end{figure}

We now estimate the number of pixels $M$ required to achieve $5$\,bpp with $P_e^{M} = 10^{-3}$ using a coherent-state transmitter and the optimum code-JDR pair. We evaluate the Burnashev-Holevo lower bound to the error exponent, $E_{\rm LB}(N_S,R) \le E(N_S,R)$, for the states $\left\{|\sqrt{N_S}\rangle, |-\sqrt{N_S}\rangle\right\}$, for which $\langle -\sqrt{N_S}|\sqrt{N_S}\rangle = e^{-2N_S}$ (Section 4 of Ref.~\cite{Bur98}). Figure~\ref{fig:EE_CSBPSK} shows contours of constant $M_{\rm UB} \equiv -\ln\epsilon/E_{\rm LB}(N_S,R)$ in the PIE ($R/N_S$) vs. $N_S$ plane for $\epsilon = 10^{-3}$. At $R/N_S = 5$\,bpp, we find that $M_{\rm UB} = 4800$. Therefore, in order to attain $5$ bpp at $P_e^{(M)}\le 10^{-3}$, the minimum number of pixels required satisfies $M \le 4800$. Given that the rate is about $2/3$ of capacity at the point where $5$\,bpp is barely reached (see the dashed magenta lines in Fig.~\ref{fig:EE_CSBPSK}), the random-coding bound is likely to be a fairly good estimate of the actual number of pixels required for an optimal code-JDR pair. A tighter upper bound and a tight lower bound on $M$ may be obtainable by using the recent results on the second-order asymptotic analysis of the quantum relative entropy~\cite{Li12, Tom12}.

\subsection{Coherent-state probe: Hadamard code, Green Machine JDR}
The probability of word error for this probe-code-JDR combination is given by the probability of erasure times the probability the erasure is mapped to an incorrect codeword, $P_e^{(M)} = (M-1)e^{-MN_S}/M$. It follows that the PIE obeys $C(N_S)/N_S = {\log_2(M)\left(1-e^{-MN_S}\right)}/{MN_S}$\,bpp, from which it is easy to deduce that achieving $5$\,bpp at $P_e^{(M)} \le 10^{-3}$, will require $M \approx 2^{35}$ pixels.  This pixel number is many orders of magnitude higher than what is required by the optimal code-JDR pair.

\subsection{W-state probe: Hadamard code, JDR}
The W-state transmitter, along with a Hadamard binary-phase code and the JDR shown in Fig.~\ref{fig:Wstate_transmitter}, can read $\log_2M$ bits using one transmitted photon at $P_e^{(M)}=0$. Therefore, to achieve $5$\,bpp, at $P_e^{(M)} \le 10^{-3}$, an $M = 32$ pixel memory suffices. This demonstrates the {\em huge} error-exponent benefit enjoyed by a quantum (spatially-entangled) transmitter in comparison with the coherent-state probe---even when the as yet unknown optimal JDR for a capacity-achieving coherent-state code may become available.

\section{Conclusions and Discussion}

We showed that using a coherent-state probe, on-off amplitude modulation, and signal-shot-noise-limited direct detection (a highly optimistic model for conventional CD/DVD drives), one cannot read any more than about $0.5$ bits per transmitted photon. We then showed that a coherent-state transmitter, in conjunction with a binary-phase-shift-keyed encoding, can read an unlimited number of bits reliably per expended photon, if non-standard joint-detection measurements are allowed at the receiver. This capacity performance of coherent states approaches the Holevo bound to capacity in the high photon-information-efficiency low-photon-flux regime. However, with a coherent-state source and binary phase encoding, if the receiver is constrained to detect the reflected light from each memory pixel one at a time followed by classical signal processing---all conventional optical receivers fall in this category---then the highest photon efficiency achievable caps off at about $2.89$ bits per photon. Thus, joint detection receivers are needed to bridge the gap to the Holevo capacity, which allows for unbounded photon efficiency for optical reading. We exhibited one example of a BPSK code-JDR pair that can bridge part of that gap, and attain an unbounded PIE. However, this example has a poor error-exponent performance. In particular, in order to attain $5$\,bpp at a word-error probability $P_e^{M} \le 10^{-3}$, it requires coding over $M \approx 2^{35}$ memory pixels, unlike the $M \approx 4800$ pixels required by the unknown optimal code-JDR pair to attain $5$\,bpp and $P_e^{M} \le 10^{-3}$ with a coherent-state probe and BPSK modulation. Finally, we showed that a non-classical W-state probe can read $\log_2(M)$ bits of data using a single photon in an $M$-mode spatially-entangled uniform-superposition state.  That performance is realized with a BPSK Hadamard code and a structured interferometric receiver which uses a linear-optical circuit of beam splitters and single-photon detectors. It attains $5$\,bpp and $P_e^{M}  = 0$ with just $M=32$ pixels, demonstrating the huge error-exponent advantage afforded by a quantum transmitter state.

\begin{figure}
\centering
\includegraphics[width=0.7\columnwidth]{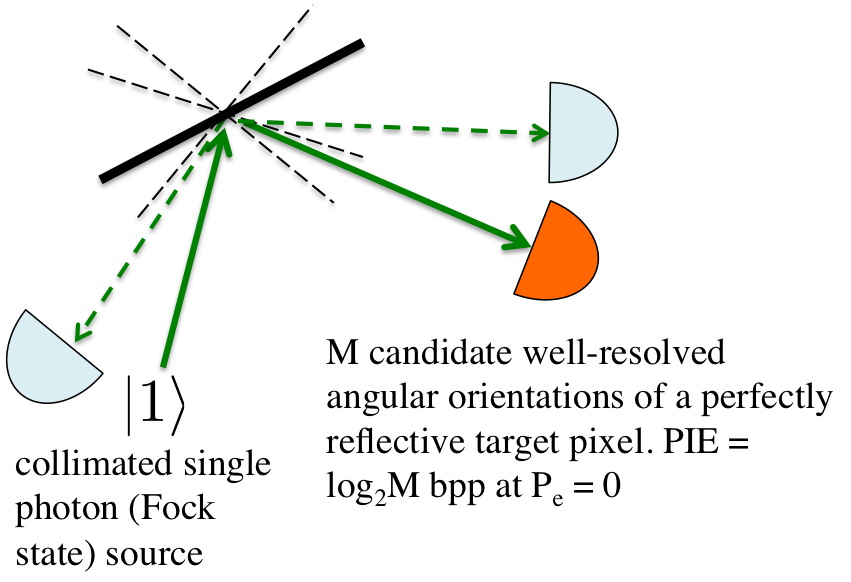}
\caption{$M$ angular positions of a perfectly-reflective mirror can encode $\log_2M$ bits of information that can be read error-free, in principle, by a well-collimated single-photon Fock state source, and an array of unity-detection-efficiency single-photon detectors.}
\label{fig:rotating_mirror}
\end{figure}

That the W-state transmitter can read any number of bits of information using just one photon should come as no surprise. Consider the thought experiment shown in Fig.~\ref{fig:rotating_mirror}. A perfectly-reflective mirror encodes information using $M$ well-resolved angular orientations, such that a well-collimated single-photon beam, reflected by the mirror, is detected via an array of single-photon detectors, one matched to each of the mirror's angular positions. Like the W-state transceiver, this arrangement reads $\log_2(M)$ bits of information using one probe photon. These angular well-resolved orthogonal modes can be replaced by any set of $M$ orthogonal space-time-polarization modes of light that the target (memory) can excite using the incident single-photon state. For the W-state example, the orthogonal modes that the spatially-entangled photon excites are a set of spatially-overlapping mutually-orthogonal spatial modes corresponding to the binary Hadamard codewords. This is quite similar to the mutually-orthogonal chip waveforms of a spread-spectrum code-division multiple-access (CDMA) system.

Before concluding, it is in order to comment briefly on the effect that loss has on optical reading. In a practical setting, loss would be incurred at various points in the reading setup: at the transmitter that generates the probe light, in the transmission to the memory pixel, in absorption and scattering of the probe light at the pixel, in collecting the reflected light at the receiver, and in the sub-unity quantum efficiency of the single-photon detector. It turns out that the capacities of all the coherent-state systems we have considered in this paper degrade gracefully with loss, i.e., these capacities have the same formulas as given in the paper for no loss with the average number of transmitted photons $N_S$ replaced by the average number of detected photons $\kappa N_S$, where $1 - \kappa \in (0, 1]$ is the end-to-end loss. Furthermore, the number of pixels $M$ needed to get a desired number of bits per {\em detected} photon for a given word-error probability does not change from what we found for the lossless case for bits per {\em transmitted} photon. For example, consider the binary-phase Hadamard code and a coherent-state probe, with the Green Machine JDR.   The resulting word-error probability in the lossy scenario is $P_e^{(M)} = (M-1)e^{-\kappa M N_S}/M \approx e^{-\kappa M N_S}$ for $M \gg 1$. So, $P_e^{(M)} = 10^{-3}$ requires $\kappa M N_S \approx 7$ photons, and since ${\rm PIE} \approx \log_2(M)/\kappa M N_S$ bits per {\em detected} photon (bpdp), attaining $5$\,bpdp requires $M \approx 2^{5 \times 7} = 2^{35}$ pixels. The {\em single-shot} W-state transceiver is able to attain $P_e^{(M)} = 0$ for the binary-phase Hadamard code and ${\rm PIE}= \log_2(M)$\,bpp using one photon in the lossless case, thereby attaining $5$\,bpp using $M = 2^5 = 32$ pixels. However, the performance of the W-state transceiver degrades rapidly with loss, smoothly transitioning to that of the coherent-state and Green Machine JDR for high loss. The W-state has an erasure probability equal to $1-\kappa$. Assuming we send $K$ copies of the W state towards the same set of Hadamard-coded pixels, and that we randomly assign a codeword to every erasure, we are left with a word-error probability $P_e^{(M)} = (M-1)(1 - \kappa)^K/M \approx e^{-\kappa K}$, for $M \gg 1$, $K \gg 1$. PIE in bits per detected photons is then ${\rm PIE} \approx \log_2(M)/\kappa K$. Thus, to get to $5$\,bpdp with $P_e^{(M)} = 10^{-3}$ when $\kappa \ll 1$ we need $M \approx 2^{35}$ pixels, just as we have for the coherent-state probe. For a single-shot W-state transmission, attaining $P_e^{(M)} = 10^{-3}$ requires $\kappa \ge 0.999$, an extraordinarily demanding task considering that single-photon detectors with 99.9\% quantum efficiency have yet to be built and there are many other sources of loss in the optical reading setup.

Finally, an interesting thing to note is that our W-state system is a special case of the Aaronson-Arkhipov (AA) {\em boson-sampling} model~\cite{Aar11}, which inputs $\sim$$\sqrt{M}$ single photons in independent spatial modes into an $M$-mode passive linear-optic circuit implementing a unitary mode transformation ${\hat {\boldsymbol a}}_{\rm out} = U{\hat {\boldsymbol a}}_{\rm in}$, followed by ideal photon counting at that transformation's $M$ output ports. The AA model was shown to be able to efficiently solve a sampling problem---that of sampling from a probability distribution comprised of the permanents of a set of matrices derived from $U$---a problem believed to be classically hard. The AA model is, however, not known to subsume universal quantum computation. A recent paper reported evidence supporting the proposition that a lossy AA system, or an AA system with mode mismatch in the linear-optic mode transformation, are likely to be classically hard to simulate, and might thereby retain some of their quantum power~\cite{Roh12}. This leads us to speculate that a multi-photon multimode transmitter could be more resilient to loss than the W-state, and hence outperform coherent states in optical reading even in the presence of loss.

Reference~\cite{Guh12a} will address the capacity of optical reading at all values of the probe photon number constraint $N_S$. There we will show that not only do quantum probes achieve a higher error exponent, they can get a fundamentally higher capacity in the high spectral-efficiency $(N_S \gg 1)$ regime. In~\cite{Guh12a} we will also consider the capacity of assisted reading, i.e., when the transmitter retains idler modes $\{{\hat a}_I^{(m,k)}\}$ at the transmitter that are entangled with the signal modes $\{{\hat a}_S^{(m,k)}\}$ which were sent towards the pixel, and joint detection is performed over the retained and returned modes. This arrangement is then the optical-reading version of quantum illumination, which has been previously studied for target detection~\cite{Tan08} and eavesdropping-immune communication~\cite{Shapiro09}.  The ultimate capacity and error-exponent performance of multi-mode transmitters remain subjects of ongoing work.

\section{Acknowledgements}

The authors thank Ranjith Nair, Brent J. Yen, Zachary Dutton, Mark M. Wilde, Stefano Pirandola and Jonathan L. Habif for useful discussions. The authors thank Marco Dalai,  Stefano Pirandola, and Peter P. Rohde for useful comments on a draft. SG was supported by the DARPA Information in a Photon (InPho) program under contract number HR0011-10-C-0162. The views and conclusions contained in this document are those of the authors and should not be interpreted as representing the official policies, either expressly or implied, of the Defense Advanced Research Projects Agency or the U.S. Government. JHS acknowledges support from the ONR Basic Research Challenge Program.


\end{document}